\documentclass[aps,prl,reprint]{revtex4-1}
\pdfoutput=1
\usepackage[pdftex]{graphicx}
\usepackage{dcolumn}   % needed for some tables
\usepackage{bm}        % for math
\usepackage{amssymb}   % for math
\usepackage{ifpdf}

\begin{document}

\title{A low-dimensional model predicting geometry-dependent dynamics of large-scale coherent structures in turbulence}

\author{Kunlun Bai}
\affiliation{Department of Mechanical Engineering and Materials Science, Yale University, New Haven, CT 06511, USA}

\author{Dandan Ji}
\affiliation{Department of Mechanical Engineering and Materials Science, Yale University, New Haven, CT 06511, USA}

\author{Eric Brown}
\affiliation{Department of Mechanical Engineering and Materials Science, Yale University, New Haven, CT 06511, USA}
\email{eric.brown@yale.edu}

%\date{\today}

\begin{abstract}
      We test the ability of a general low-dimensional model for turbulence to predict geometry-dependent dynamics of large-scale coherent structures, such as convection rolls. The model consists of stochastic ordinary differential equations, which are derived as a function of boundary geometry from the Navier-Stokes equations \cite{brown_model_2008, brown_azimuthal_2008}.  We test the model using Rayleigh-B\'enard convection experiments in a cubic container. The model predicts a new mode in which the alignment of a convection roll switches between diagonals. We observe this mode with a measured switching rate within 30\%  of the prediction.
\end{abstract}

\maketitle     
    
%%% Introduction
      Large-scale coherent flow structures in turbulence -- such as convection rolls in the atmosphere -- are ubiquitous and can play a dominant role in heat and mass transport. A particular challenge is to predict dynamical states and their change  with different boundary  geometries,  for example in the way that local weather patterns depend on the topography of the Earth's surface.  However, the Navier-Stokes equations that describe flows are impractically difficult to solve for turbulent flows, so low-dimensional models are desired. 

      It  has long been  recognized that the dynamical states of large-scale coherent structures are similar to those of low-dimensional dynamical systems  models \cite{lorenz_deterministic_1963} and stochastic ordinary differential equations   \cite{brown_large-scale_2007, de_la_torre_slow_2007, thual_stochastic_2014, rigas_stability_2015}.  However, such models tend to be descriptive rather than predictive, as parameters are typically fit to observations, rather than derived \citep{HolmesLumley1996}. In particular, dynamical systems models tend to fail at  quantitative predictions of new dynamical states in regimes outside where they were parameterized. In this letter we demonstrate a proof-of-principle that a general low dimensional model can quantitatively predict the different dynamical states of large-scale coherent structures in different geometries. 

\begin{figure}[]
\includegraphics[width=0.4\textwidth]{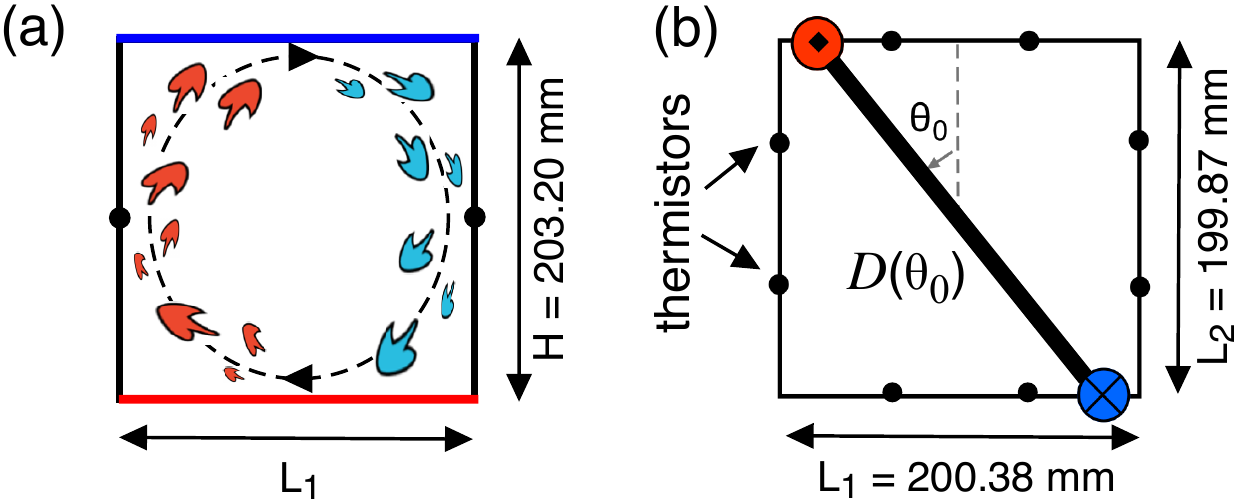}
\caption{(a) A  side view of the large-scale circulation (LSC), indicated by the dashed line. Hot and cold features are red and blue,  respectively (online). (b) Top view of a horizontal cross-section  at mid-height of the cubic container.  Thermistor locations on the side wall are indicated by small circles. The orientation of the LSC is defined as the angle $\theta_0$ between the hot side of the circulation plane (thick solid line) and the vertical dashed line. The length of the circulation plane across a horizontal cross-section $D(\theta_0)$ determines the model potential.}
\label{fig_sketch_cell}
\end{figure}

      The model system is Rayleigh-B\'enard convection, in which a  fluid is heated from  below and cooled from above to generate buoyancy-driven convection  \cite{ahlers_heat_2009, lohse_smallscale_2010}. This system exhibits robust large-scale coherent structures that retain the same organized flow structure over long times. For example, in upright cylindrical containers  of aspect ratio 1, a large-scale circulation (LSC) forms. This LSC consists of  temperature and velocity fluctuations which, when coarse-grain averaged, collectively form a single convection roll in a vertical plane \cite{Krishnamurti_1981},  as shown in Fig.~\ref{fig_sketch_cell}a. Various dynamics of the LSC have been reported, including spontaneous meandering of the orientation $\theta_0$ in a horizontal plane, and an advected oscillation which appears as a torsional or sloshing mode \cite{ciliberto_large-scale_1996, brown_rotations_2006, xi_azimuthal_2006, brown_rotations_2006, xi_cessations_2007, funfschilling_plume_2004, brown_origin_2009, xi_origin_2009}. As an example of different dynamical states in different geometries, if instead the axis of the cylinder is aligned horizontally, $\theta_0$ tends to align with the longest diagonals of the cell, and  oscillates periodically between diagonals and around individual corners \cite{song_dynamics_2014}.    

      While there are several low-dimensional models for LSC dynamics \cite{sreenivasan_mean_2002, benzi_flow_2005, araujo_wind_2005, resagk_oscillations_2006}, only one by Brown \& Ahlers has  made predictions dependent on container geometry \cite{brown_large-scale_2007, brown_model_2008, brown_azimuthal_2008}. The model consists of a pair of stochastic ordinary differential equations,  using the empirically known,  robust LSC structure as an approximate solution to the Navier-Stokes equations.  The resulting dynamical equation for $\theta_0$  is

\begin{equation}
\label{eq_theta_eq}
\ddot{\theta}_0 = - \frac{\dot{\theta}_0\delta}{\tau_{\dot{\theta}}\delta_0} - \nabla V_g(\theta_0) + f_{\dot{\theta}}(t) \ . 
\end{equation}   

\noindent     The first term on the right is a damping term where   $\tau_{\dot{\theta}}$ is a damping time scale.  A  separate  stochastic ordinary differential equation describes the fluctuations of $\delta$  around its stable fixed point $\delta_0$ \citep{brown_model_2008}. $f_{\dot{\theta}}$ is a stochastic forcing term representing the effect of small-scale turbulent fluctuations and  is modeled as Gaussian white noise with diffusivity $D_{\dot\theta}$.  This model is mathematically equivalent to diffusion in a potential landscape $V_g(\theta_0)$. The potential $V_g$ represents the pressure of the sidewalls acting on the LSC, and is given by

\begin{equation}
\label{eq_potential}
V_g(\theta_0) = \left<\frac{ 3\omega_{\phi}^2 H^2 }{4 D(\theta_0)^2}\right>_{\gamma}
\end{equation} 

\noindent where $\omega_{\phi}$ is the turnover frequency of the LSC, and $H$ is the height of the container \citep{brown_azimuthal_2008}.  This includes an  update to \citep{brown_azimuthal_2008} of the numerical coefficient for aspect ratio 1  containers \citep{song_dynamics_2014}.  The  notation $\langle ...\rangle_{\gamma}$  represents a smoothing of the potential over the width $\gamma=\pi/10$ of the LSC \citep{song_dynamics_2014}. $D(\theta_0)$ is the distance across a horizontal cross-section of the cell, as a function of $\theta_0$,  illustrated in Fig.~\ref{fig_sketch_cell}b.  Thus, $D(\theta_0)$, and consequently $V_g$ and Eq.~\ref{eq_theta_eq}  can be predicted explicitly for any system geometry, with the caveat that in this form of the model the geometry must support a single-roll LSC. 
 
\begin{figure}[]
\includegraphics[width=0.45\textwidth]{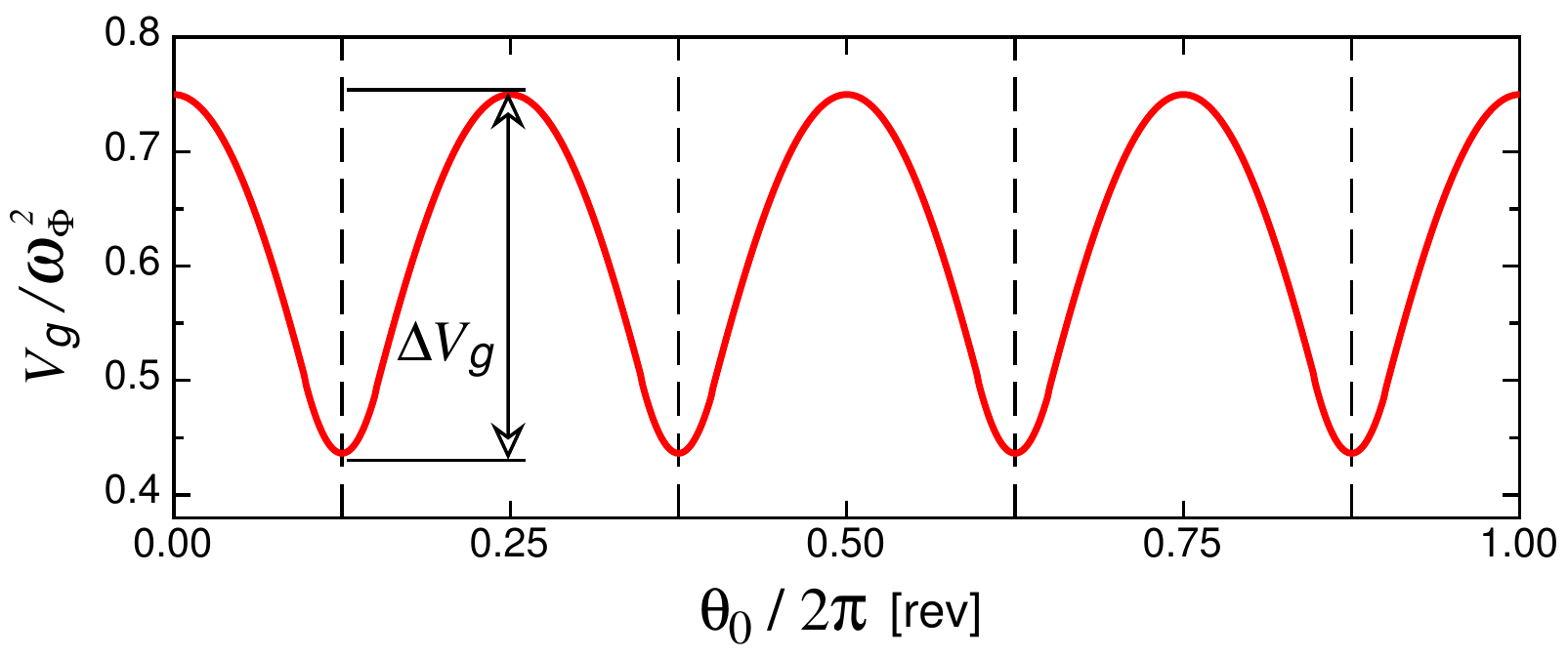}
\caption{The model potential $V_g(\theta_0)$ for a cubic cell (Eq.~\ref{eq_potential}). The vertical dashed lines indicate the location of the four corners  where the potential minima occur. Eq. \ref{eq_theta_eq} describes diffusive fluctuations of $\theta_0$ in this potential, which can occasionally cross the barriers $\Delta V_g$  to switch between corners. 
}
\label{fig_potential}
\end{figure}

      This model  and its extensions have successfully described all of the known dynamics of the LSC \cite{brown_large-scale_2007, brown_model_2008, brown_azimuthal_2008, brown_origin_2009, brown_Coriolis_2006, song_dynamics_2014, zhong_dynamics_2015}. Since the model is derived from first principles, the model terms can be predicted and are typically accurate within a factor of 2.  The only  required fit parameter is $D_{\dot\theta}$ which can be fit to independent measurements \cite{brown_model_2008}. The model has described dynamics dependent on the geometric potential $V_g$ \citep{song_dynamics_2014}, although in that case a correction  was made to $V_g$  for the nonzero width of the LSC, and  another parameter was fit  to better describe data. Since the model was adjusted to describe results after they were observed, it  has not yet been shown that the model can predict geometry-dependent dynamics before their observation.

      In this letter, we test the model prediction of the existence of  a previously unobserved mode: a stochastic switching of $\theta_0$ between potential wells \cite{brown_azimuthal_2008}. We test this prediction in a cubic container which has 4 potential wells and 4 potential barriers of equal height, calculated from Eq.~\ref{eq_potential}, and shown in Fig.~\ref{fig_potential}.   The cubic geometry prevents a competing periodic oscillation mode, which could occur  if  one potential barrier is smaller such that the system could oscillate in the  wider well  surrounding two corners \cite{song_dynamics_2014}. This is the first example of testing a quantitative prediction of a  previously unobserved geometry-dependent mode of the LSC, and without any flexibility  or free parameters in the model.

%%% Experimental setup
      The cubic container is based on the design of \citep{brown_finiteConductivity_2005}. It has dimensions $H=203.20$ $mm$, $L_1=200.38$ $mm$, and $L_2=199.87$ $mm$, illustrated in Fig.~\ref{fig_sketch_cell}. The variation of the cell dimensions due to bowing of the sidewall, epoxy to seal gaps and cover thermistors, and holes for filling water are each less than 0.7 $mm$. The cell is filled with degassed and deionized water at mean temperature 23.0 $^oC$, for a Prandtl number $Pr\equiv\nu/\kappa=6.4$ ($\kappa$ is the thermal diffusivity, and $\nu$ is the kinematic viscosity). We report measurements at Rayleigh number $Ra\equiv\alpha g \Delta T H^3/\kappa\nu=4.8\times10^8$ ($\Delta T=3.8$ $^oC$ is the temperature difference between top and bottom plate, $\alpha$ is the isobaric thermal expansion coefficient, and $g$ is the acceleration of gravity). The averaged standard deviation of the plate-temperature variation in space and time is 0.005$\Delta T$. The cell is isolated from room temperature variations as in \citep{brown_finiteConductivity_2005}. It is leveled within 0.03 degree.
           
      Fluid  temperature is recorded by thermistors  placed  in blind holes in the  acrylic sidewall,  within 0.5 $mm$ of the fluid surface \cite{brown_rotations_2006}. Thermistor  locations are  equally spaced in angle $\theta$ in the horizontal plane at mid-height of the container as shown in Fig.~\ref{fig_sketch_cell}b, such that the four corners are located at $\theta = \frac{1}{8}$, $\frac{3}{8}$, $\frac{5}{8}$, and $\frac{7}{8}$ $rev$. The relative error on thermistor  measurements is 2.5 $mK$,  determined from calibrations.  The LSC can be detected  by the hot fluid it  pulls up on one side and the cold fluid  it pulls down the other side, as shown in Fig. \ref{fig_sketch_cell}a. The thermistor temperatures $T$  are fit by $T = T_0 + \delta cos(\theta-\theta_0)$  to obtain the LSC orientation $\theta_0$, and half the horizontal temperature difference $\delta$ which characterizes the strength of the LSC, as in \cite{brown_rotations_2006}.

%%% Results
\begin{figure}[]
\includegraphics[width=0.48\textwidth]{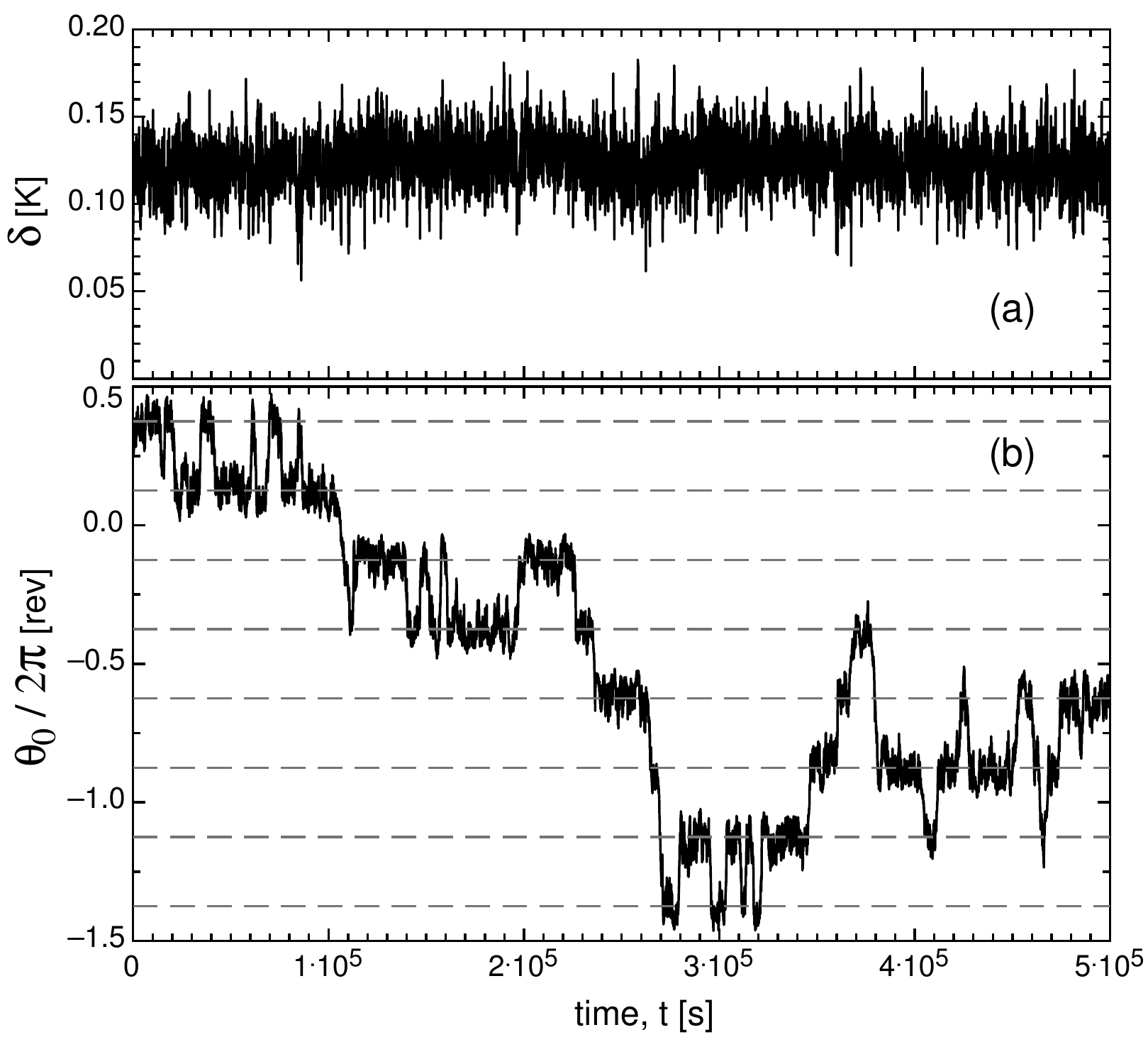}
\caption{Typical time series of the strength $\delta$ and the orientation $\theta_0$ of the LSC in (a) and (b), respectively. The horizontal dashed lines in (b) indicate the locations of the four corners in the cubic container. Stochastic switching of $\theta_0$ between corners is observed, as predicted \citep{brown_model_2008}.}
\label{fig_time_series}
\end{figure}

      A typical time series of  the strength $\delta$ and orientation $\theta_0$ of the LSC is shown in Fig.~\ref{fig_time_series}.  $\theta_0$ meanders erratically as in cylindrical containers \cite{brown_rotations_2006, xi_azimuthal_2006, song_dynamics_2014}. $\theta_0$ also prefers to align with the corners (dashed lines in Fig.~\ref{fig_time_series}b), which is different  from upright cylindrical containers, and similar to previous measurements in rectangular containers \citep{daya_ecke_2001, zhou_oscillation_rectangularCell_2007} and horizontal cylinders \cite{song_dynamics_2014}. Such preference is expected since corners correspond to potential minima (Fig.~\ref{fig_potential}). Finally, $\theta_0$ switches between corners,  apparently randomly. In previous studies it was found that $\theta_0$ could reorient quickly due to cessation  and reformation of the LSC,  which is characterized by a drop of  the LSC strength $\delta$  to effectively zero \cite{brown_reorientation_2005}.  In the present study, $\delta$ fluctuates around its stable fixed point value $\delta_0=0.124$ $K$ without dropping below $0.46\delta_0$, which indicates the  switching observed here occurs without cessation. We also observe that the LSC samples all four corners, not just oscillating back and forth between two corners as observed by Song et al. \cite{song_dynamics_2014}. These  qualitative observations are all consistent with the model prediction of stochastic switching across potential barriers.

\begin{figure}[]
\includegraphics[width=0.47\textwidth]{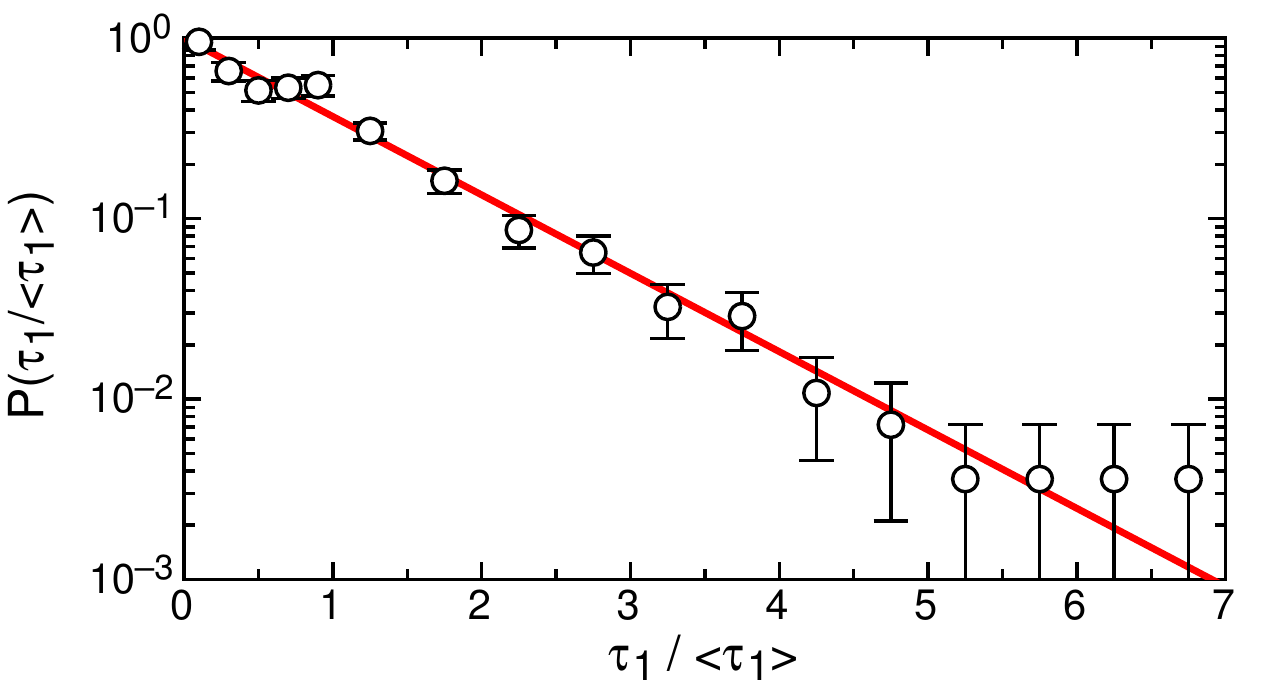}
\caption{The probability distribution $P(\tau_1/\langle\tau_1\rangle)$ of the time intervals between switching of $\theta_0$ from one corner to another. Solid line: the function $P(\tau_1/\langle\tau_1\rangle)=\exp(-\tau_1/\langle\tau_1\rangle)$ representing  a Poisson (random) distribution.}
\label{fig_time_interval}
\end{figure}

      To  characterize the randomness of the switching, we measure the distribution of the time intervals $\tau_1$ between switching events.  For the purposes of counting events, the LSC  is counted to be in one corner until it crosses  all the way to the orientation of  an adjacent corner.  This method avoids counting  extraneous events due to the jitter of $\theta_0$ around the peaks of the potential. The probability distribution $P(\tau_1/\langle\tau_1\rangle)$ is shown in Fig.~\ref{fig_time_interval}, where $\langle\tau_1\rangle$ is the average time interval between switching. The  fractional error  on each point is equal to the inverse square root of the number of events in each bin. Notably, there is no peak for $\tau_1 > 0$,  confirming that the switching is not periodic as observed in Song et al. \citep{song_dynamics_2014}.  The data are  consistent with the exponential function $P(\tau_1/\langle\tau_1\rangle)=exp(-\tau_1/\langle\tau_1\rangle)$ shown  as the line in Fig.~\ref{fig_time_interval}, which represents Poisson  statistics, i.e. randomly distributed  events in time, as predicted for the model of  overdamped diffusion  across a potential barrier \citep{brown_azimuthal_2008}. 

      For a quantitative prediction, the rate of  switching between corners can be modeled as a  fluctuation-driven crossing of a potential barrier. This was done previously \citep{brown_model_2008} by  simplifying Eq.~\ref{eq_theta_eq}  to the one solved by Kramers \cite{kramers_brownian_1940} by approximating  $\delta=\delta_0$,  which is valid if the fluctuations of $\delta$ around its stable fixed point $\delta_0$  are small. In the overdamped limit, the  number of switching events per unit time is given by

\begin{equation}
\label{eq_omega_of_delta_model}
\omega =  \frac{ \sqrt{c_{min}c_{max}}\tau_{\dot{\theta}} }{2\pi} exp \left( - \frac{ \Delta V_g }{D_{\dot{\theta}} \tau_{\dot{\theta}}} \right).
\label{eqn_switching_rate}
\end{equation}   

\noindent $c_{min}=15\omega_{\phi}^2/\pi$ and $c_{max}=3\omega_{\phi}^2/2$ are the curvatures $|d^2V_g/d\theta^2|$ at the minimum and maximum of the potential, respectively. The potential barrier $\Delta V_g = \frac{3}{8}(1-\frac{\gamma}{2})\omega_{\phi}^2$ is  calculated from Eq.~\ref{eq_potential}  \citep{song_dynamics_2014}.  The damping time scale $\tau_{\dot{\theta}}$=17.5 $\pm$ 0.5 $s$ and the diffusivity $D_{\dot\theta}=(2.37 \pm 0.07) \times 10^{-6}$ $rad^2/s^{3}$ are fitted independently from the mean-square change in $\dot{\theta}_0$ over time as in \citep{brown_model_2008}. The circulation rate  $\omega_{\phi}$=0.022 $\pm$ 0.003 $s^{-1}$ is obtained  by first calculating the speed of the LSC as the distance $H/4$ between 2 vertically separated thermistors  in the path of the LSC, divided by the time of peak correlation between their signals (16.6 $\pm$ 0.7 $s$), and further divided by the path length of the LSC, which is assumed to be between  a  rectangular path along a diagonal of length $2(1+\sqrt{2})H$  and a nearly ellipsoidal path  of length $\pi(1+\sqrt{2})H/2$.  With these parameter values and Eq.~\ref{eqn_switching_rate},  the predicted  switching rate $\omega=(0.9 \pm 0.6)\times 10^{-4}$ $s^{-1}$. This prediction is  smaller than the measured switching rate $\bar{\omega} = 1.3 \times 10^{-4}$ $s^{-1}$ (251 events measured over 21.7 days) by 40\%, while consistent within error. 
      
      Alternatively, we can predict  the parameter value $\tau_{\dot{\theta}}=26.9$ $s$ from first principles \citep{brown_model_2008}. This  value is higher than the  independently measured value by 54\%, increasing the predicted $\omega$ by  460\%. This  example indicates that the prediction of $\omega$ is very sensitive to parameter values, due to the exponential term in Eq.~\ref{eqn_switching_rate}. This sensitivity means that the agreement within 40\% for $\omega$ implies much better accuracy of 9\% for individual model parameters. For our variation of cell dimensions of 0.7 $mm$ (0.35\%), $\Delta V_g$ could change by 0.95\%, causing the  predicted $\omega$ to change by 3.5\%.  This confirms our cell is still uniform enough to compare to predictions for a cubic cell. 
      
      To provide a  stricter test of the model, we extend the prediction of switching rate $\omega$   to be a function of $\delta$ while still using the dynamics of $\delta$ from that original model.   In principle, the fluctuations  of $\delta$  around the stable fixed point $\delta_0$ can affect both the damping and potential terms in Eq.~\ref{eq_theta_eq}.  To account for this,  we remove the model approximation of a fixed $\delta=\delta_0$  used in the original calculation of $\omega$ (Eq.~\ref{eqn_switching_rate})  \cite{brown_model_2008}.   We can explicitly write the $\delta$-dependence into the model since $\delta$ varies slowly, i.e. the timescale $\tau_{\delta}$ that governs $\delta$ is much larger than the timescale $\tau_{\dot{\theta}}$ that governs $\theta_0$ \citep{brown_model_2008}. Thus, the damping timescale $\tau_{\dot\theta}$ in Eq.~\ref{eqn_switching_rate} can be replaced with  $\tau_{\dot\theta}\delta_0 / \delta$ as in Eq.~\ref{eq_theta_eq}. In addition, since $\omega_{\phi}$ was  assumed to be proportional to $\delta$ in the original model \cite{brown_model_2008},  but Eq.~\ref{eq_potential} was originally written  with the implicit approximation $\delta = \delta_0$, $\Delta V_g$ can be  generalized to $\Delta V_g(\delta) = \frac{3}{8}\left(1-\frac{\gamma}{2}\right) \left(\frac{\omega_{\phi}\delta}{\delta_0}\right)^2$. Using the same  overdamped Kramers solution for the barrier crossing problem as in Eq.~\ref{eqn_switching_rate}, the switching rate becomes
 
 \begin{equation}
\label{eq_omega_of_delta_model}
\omega(\delta) =\frac{ \sqrt{c_{min}c_{max}}\tau_{\dot{\theta}} \delta_0 }{2\pi\delta} \exp\left( - \frac{ 3\omega_{\phi}^2 \delta^3}{8D_{\dot{\theta}} \tau_{\dot{\theta}}\delta_0^3} \left(1-\frac{\gamma}{2}\right) \right) \ .
\label{eqn_switching_rate_delta}
\end{equation}   

\noindent   This expression represents the rate of switching per unit time at  each value of $\delta$.

      To compare this prediction with  measurements, we  calculate the corresponding measured value of $\omega(\delta)$ from $\omega(\delta) = \bar{\omega} P_s(\delta) / P(\delta)$, where $P(\delta)$ is the probability distribution of $\delta$ during  an entire data set, and $P_s(\delta)$ is the distribution of $\delta$  during switching events.  For each switching event,  we  use the value of $\delta$  the last time that $\theta_0$ crosses the potential maximum.

\begin{figure}[]
\includegraphics[width=0.45\textwidth]{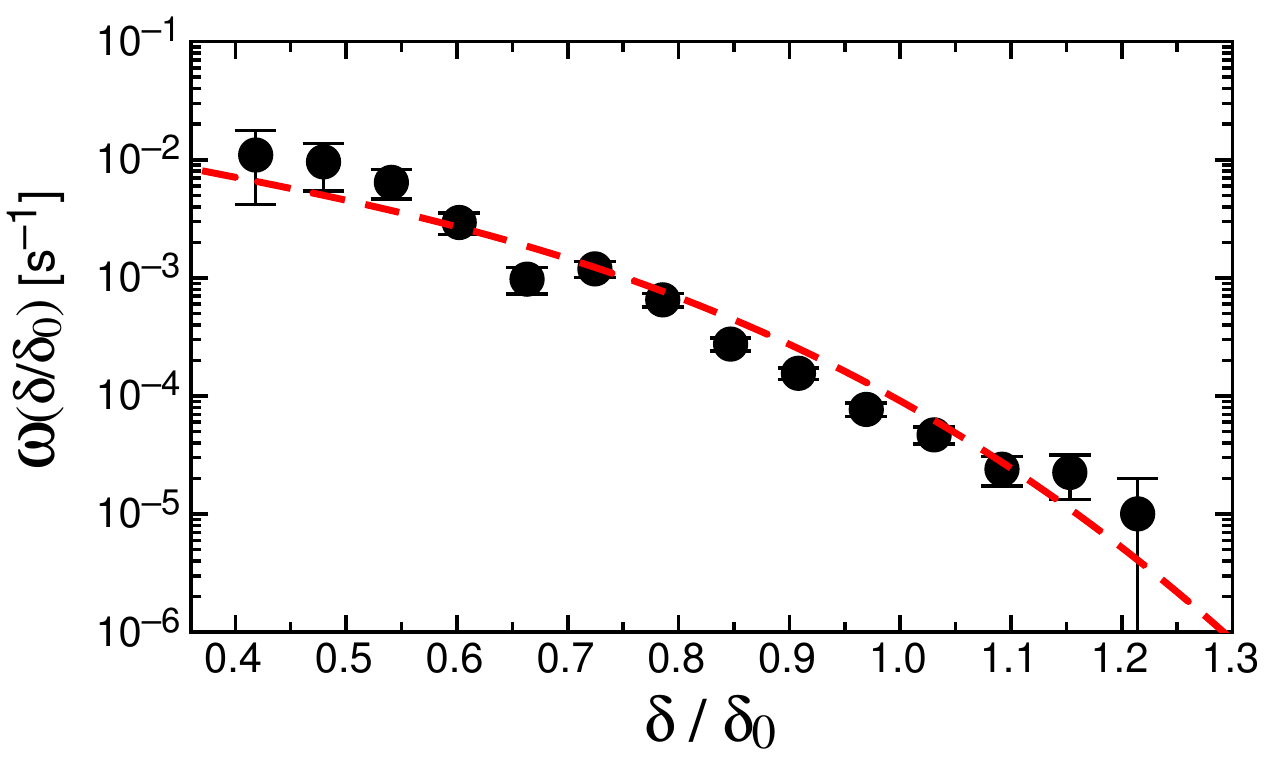}
\caption{The switching rate $\omega(\delta)$. Circles: measurements. Dashed line: model prediction from Eq.~\ref{eq_omega_of_delta_model}.
}
\label{fig_switching_rate}
\end{figure}
     
      Figure \ref{fig_switching_rate} shows a comparison of the measured $\delta$-dependent switching rate $\omega(\delta)$ and the model prediction  from Eq.~\ref{eqn_switching_rate_delta}. The  trend of the data is captured well by the model, as the root-mean-square difference between measured and predicted $\omega(\delta)$ is 50\% over 3 decades of $\omega$. The $\delta$-dependence in $\omega(\delta)$ leads to a modified prediction of the  average switching rate: $\int \omega(\delta)P(\delta) d\delta = (1.7 \pm 1.1) \times10^{-4}$ $s^{-1}$, which is consistent with, and within 30\% of the measured switching rate $\bar{\omega}=1.3\times10^{-4}$ $s^{-1}$. However, this level of accuracy in $\bar{\omega}$ is  better than we should expect, since predictions of this model are  typically only accurate within a factor of 2 or 3 due to the approximations made to obtain Eq.~\ref{eq_theta_eq} \citep{brown_model_2008},  unless model parameters  are  fit to data  in  non-independent measurements  \cite{assaf_rare_2011}.  Regardless, the agreement between the  predicted and measured $\omega(\delta)$ is  exceptionally good for a low-dimensional model, considering parameter values $\tau_{\dot\theta}$, $D_{\dot\theta}$, and $\omega_{\phi}$ are determined from independent measurements  and the geometry dependence is predicted from first principles. 
        
      The increase of  the switching rate $\omega$ as $\delta$ decreases can be understood in terms of Eqs.~\ref{eq_theta_eq} and \ref{eqn_switching_rate_delta}. Small $\delta$ means a weaker LSC which leads to both smaller damping in Eq.~\ref{eq_theta_eq} and potential barriers in Eq.~\ref{eqn_switching_rate_delta}.  Both of these effects allow fluctuations to drive the system over the potential barriers more easily,  resulting in a higher $\omega$.

%%% Conclusions
      To summarize,  we observe that LSC orientation $\theta_0$ switches between corners as a Poisson process, as predicted  \citep{brown_model_2008}. The prediction of the   average switching rate $\bar{\omega}$ is 30\%  above the measured value,  within error, while the prediction of $\omega(\delta)$  captures the trend in $\delta$  with  a root-mean-square difference of only 50\% over three decades of $\omega$ (Fig.~\ref{fig_switching_rate}). The switching can be understood as  a turbulent-fluctuation-driven crossing of a potential barrier,  where the potential is predicted from the shape of the sidewall. The switching is more likely to happen when $\delta$ is smaller, due to the decrease in both the potential barrier and  damping. 
   
      This new dynamical mode -- an examle of a dynamic that depends on geometry -- could be predicted because the low-dimensional model is derived from first principles. The key insight that allowed this derivation was  that the robustness of the LSC allows it to be plugged in as an approximate solution to the Navier Stokes equations.  The success of the prediction demonstrates that a  low-dimensional turbulence model can quantitatively predict the existence and properties of different dynamical states  and how they depend on boundary geometry.   Since this methodology can in principle be applied to other flows  dominated by large-scale coherent structures, it opens up the potential for further development of general,  low-dimensional  turbulence models.

%%% Acknowledgment 
We thank the University of California, Santa Barbara machine shop and K. Faysal for helping with construction of the experimental apparatus. This work is supported by Grant CBET-1255541 of the U.S. National Science Foundation.

\bibliographystyle{apsrev4-1} % Tell bibtex which bibliography style to use
%\bibliography{ref_KB_DJ_EB_PRL} % Tell bibtex which .bib file to use (this one is some example file in TexLive's file tree)

%merlin.mbs apsrev4-1.bst 2010-07-25 4.21a (PWD, AO, DPC) hacked
%Control: key (0)
%Control: author (72) initials jnrlst
%Control: editor formatted (1) identically to author
%Control: production of article title (-1) disabled
%Control: page (0) single
%Control: year (1) truncated
%Control: production of eprint (0) enabled
%

\end{document}